\documentstyle[prl,aps,preprint,floats]{revtex}

\begin{document}

\title{Torsion nonminimally coupled to the electromagnetic field and
birefringence}

\author{Guillermo F.~Rubilar}
\address{Departamento de F{\'{\i}}sica, Universidad de Concepci\'on,
Casilla 160-C, Concepci\'on, Chile}
\author{Yuri N.~Obukhov\footnote{On leave from: Dept. of Theoret. 
Phys., Moscow State University, 117234 Moscow, Russia} {\rm and} 
Friedrich W.~Hehl\footnote{Also at: Dept. of Phys. \& Astron.,
University of Missouri-Columbia, Columbia, MO 65211, USA}}
\address{Institute for Theoretical Physics, University of Cologne,
50923 K\"oln, Germany}


\maketitle

\begin{abstract}
In conventional Maxwell--Lorentz electrodynamics, the
propagation of light is influenced by the metric, not, however, by the
possible presence of a torsion $T$. Still the light can feel torsion 
if the latter is coupled nonminimally to the electromagnetic field $F$ 
by means of a supplementary Lagrangian of the type $\sim\ell^2 T^2\,F^2$ 
($\ell$ = coupling constant). Recently Preuss suggested a specific 
nonminimal term of this nature. We evaluate the spacetime relation of 
Preuss in the background of a general O(3)-symmetric torsion field and 
prove by specifying the optical metric of spacetime that this can yield 
birefringence in vacuum. Moreover, we show that the nonminimally
coupled homogeneous and isotropic torsion field in a Friedmann cosmos 
affects the speed of light.
\end{abstract}
\bigskip



\section{``Admissible'' couplings of torsion}

Maxwell's equations $dH=J$ and $dF=0$, with the excitation $H=({\cal
  H},{\cal D})$ and the field strength $F=(E,B)$, are generally
covariant. If supplemented by the Maxwell--Lorentz spacetime relation
$H=\lambda_0\,^\star F$ (the star denotes the Hodge operator and
$\lambda_0$ the vacuum impedance), one arrives at the standard
Maxwell--Lorentz vacuum electrodynamics and can study, for example,
the propagation of electromagentic disturbances (``light'') in vacuum.
Clearly, the metric is then the only geometrical quantity light
couples to. In particular, torsion $T^\alpha$ and nonmetricity
$Q_{\alpha\beta}$ are completely ``transparent'' to light, as has been
discussed, for example, by Puntigam et al.\ \cite{puntigam}. Light
doesn't interact with these fields under the assumptions specified.

Torsion can affect light when either the Maxwell equations $dH=J$ and 
$dF=0$ (at least one of them) or the Maxwell--Lorentz spacetime relation
$H=\lambda_0\,^\star F$ are modified. A modification of the inhomogeneous 
Maxwell equation could read $[d +\,(e_\alpha\rfloor T^\alpha)]H=J$, with 
the frame $e_\alpha$ and the interior product sign $\rfloor$; see in this
context \cite{gyrosClaus,clausannalen}. This {\em nonminimal\/} coupling 
of torsion to the Maxwell equations we call ``inadmissible'' \cite{gyros} 
since it violates the well-established conservation law of electric charge. 
The same can be said, {\em mutatis mutandis}, with respect to the 
homogeneous equation $dF=0$ and the conservation law of magnetic flux.

Therefore, we turn to the ``admissible'' nonminimal couplings that 
modify the Maxwell--Lorentz spacetime relation. One possible example 
is, see \cite{gyros}:
\begin{equation}\label{nmtorsion}
H=\lambda_0\left[1 + \gamma^2{}\,^\star\!\left(T^\alpha\wedge
T_\alpha\right)\right] \, {}^\star\!F\,.
\end{equation}
Such a dilaton type coupling preserves the light cone structure. Other 
quadratic torsion coupling are also possible, see \cite{itin} for a 
complete list. It should be understood that the Maxwell equations $dH=J$ 
and $dF=0$ are left intact, with electric charge and magnetic flux still 
conserved, but the spacetime relation (\ref{nmtorsion}) with a new constant 
of nature $\gamma$ would induce a coupling of light to torsion; here
$[\gamma] = length$. Note that the spacetime relation
(\ref{nmtorsion}) is still linear in $F$ but a quadratic torsion term
emerges; also quartic torsion terms would be possible.

\section{Preuss' coupling to a background torsion}

Recently Preuss in his thesis \cite{drpreuss} and in a talk
\cite{hannoverpreuss} suggested a different non-minimal coupling term.
His supplementary Lagrangian 4-form
\begin{equation}\label{preussnonmin}
  L_{\rm Preuss} = \lambda_0\ell^2\, ^\star \left(T_\alpha \wedge F\right) 
 \,T^\alpha \wedge F
\end{equation}
(cf. Eq.(1.59) of \cite{drpreuss}) corresponds to the spacetime
relation
\begin{equation}\label{preussnonmin1}
  H = \lambda_0\left[{}^\star F-2\ell^2\, ^\star \left(T_\alpha \wedge F
  \right) \, T^\alpha\right],
\end{equation}
since $L=-\frac{1}{2}\,H\wedge F$. The coupling constant $\ell$ has the
dimension of length. Eq.~(\ref{preussnonmin1}) should be compared with 
(\ref{nmtorsion}). Both spacetime relations are options but for parity 
reasons (\ref{preussnonmin1}) looks better. Alternatively, one could also 
modify (\ref{nmtorsion}) by inserting a star within the parenthesis: 
$\gamma^2{}\,^\star\!\left(T^\alpha\wedge T_\alpha\right)\longrightarrow 
\gamma^2{}\,^\star\!\left( T^\alpha\wedge \,^\star T_\alpha\right)$. We 
will not discuss the most general form of such quadratic torsion couplings 
here, see, however, Itin \cite{itin}. For concreteness, we confine our 
attention to the Preuss Lagrangian. Hence from now on we assume the 
spacetime relation (\ref{preussnonmin1}).

Preuss made his ansatz since he was on a search for new astronomical
tests of Einstein's equivalence principle. In addition, he took an exact 
solution found by Tresguerres \cite{TresguerresShear1,TresguerresShear2} 
within the metric-affine theory of gravity (MAG); for MAG see the review
\cite{gron96}. The main ingredient for these considerations is a
O(3)-symmetric torsion that is given by ($\vartheta^{\hat 0\hat
  1}:=\vartheta^{\hat 0}\wedge \vartheta^{\hat 1}$ etc., with
$\vartheta^\alpha$ as coframe)
\begin{equation}\label{torsion10}
  T^{\alpha}|_{{\rm O}(3)} = \pmatrix{ 
 T_{\hat 0\hat 1}{}^{\hat 0} \,{\vartheta}^{\hat 0\hat 1} \cr 
 T_{\hat 0\hat 1}{}^{\hat 1} \,{\vartheta}^{\hat 0\hat 1} \cr 
 T_{\hat 0\hat 2}{}^{\hat 2} \,{\vartheta}^{\hat 0\hat 2}
+T_{\hat 1\hat 2}{}^{\hat 2} \,{\vartheta}^{\hat 1\hat 2} \cr 
 T_{\hat 0\hat 3}{}^{\hat 3} \,{\vartheta}^{\hat 0\hat 3}   
+T_{\hat 3\hat 1}{}^{\hat 3} \,{\vartheta}^{\hat 3\hat 1}}={\frac 1 \ell}
\pmatrix{f \,{\vartheta}^{\hat 0\hat 1} \cr 
-h\,{\vartheta}^{\hat 0\hat 1} \cr 
-k \,{\vartheta}^{\hat 0\hat 2}
+g \,{\vartheta}^{\hat 1\hat 2} \cr 
-k \,{\vartheta}^{\hat 0\hat 3}   
-g \,{\vartheta}^{\hat 3\hat 1}} \,.
\end{equation} 
We introduced an overall $\ell^{-1}$ factor in (\ref{torsion10}) 
in order to keep the functions $f, g, h, k$ dimensionless. 
For the torsion with its 24 components only 4 functions are left 
open in the case of O(3)-symmetry. Preuss took $f(r), g(r),h(r),k(r)$ 
from \cite{TresguerresShear1,TresguerresShear2} and computed the
corresponding birefrigence by using a formalism of Haugan and
Kauffmann \cite{Haugan95} (see also Gabriel et al.\ \cite{Haugan91}).

\section{Birefringence for an arbitrary spherically symmetric torsion}

We have shown in the past (see \cite{Birk,drguillermo}) how one can
determine light propagation in an arbitrary spacetime by means of a
generalized Fresnel equation provided a {\em linear\/} spacetime
relation $H=\kappa(F)$, or in components $H_{\alpha\beta}=
\frac{1}{2}\, \kappa_{\alpha\beta} {}^{\gamma\delta}\,
F_{\gamma\delta}$, is specified. Recently, we applied this method also
to nonlinear electrodynamics, see \cite{Obukhov:2002xa}. Here we
demonstrate our procedure for the spherically symmetric torsion case,
but an extension to axial symmetry is straightforward. In this way we
confirm Preuss' results in a more general framework and in a very
direct manner.

\subsection{Modified constitutive tensor density of spacetime}

We start by putting (\ref{preussnonmin1}) into component form. Field
strength and torsion decompose as $F=F_{\alpha\beta}\,
\vartheta^{\alpha\beta}/2$ and $T^\alpha=T_{\beta\gamma}
{}^\alpha\,\vartheta^{\beta\gamma}/2$, respectively. Then
(\ref{preussnonmin1}) reads
\begin{equation}\label{exex}
 H= \frac{1}{2}\,H_{\alpha\beta}\, \vartheta^{\alpha\beta}= 
  {\frac{\lambda_0}{2}}\left[F_{\alpha\beta}\,^\star\vartheta^{\alpha\beta}
  -{\ell^2}\,^\star(T_\alpha\wedge \vartheta^{\beta\gamma})
  \,T_{\mu\nu}{}^\alpha\, F_{\beta\gamma}\, \vartheta^{\mu\nu}\right].
\end{equation}
Because of $ ^\star\vartheta^{\alpha\beta}=\eta^{\alpha\beta}
{}_{\gamma\delta}\,\vartheta^{\gamma\delta}/2\,$, here $\eta^{\alpha
  \beta \gamma\delta}$ is the totally antisymmetric unit tensor, we find
\begin{equation}
\kappa_{\gamma\delta}{} ^{\alpha\beta}= \lambda_0\left[\eta^{\alpha\beta}
  {}_{\gamma\delta}-2\ell^2\,^\star(T_\mu\wedge \vartheta^{\alpha\beta})
  \,T_{\gamma\delta}{}^\mu\right].
\end{equation}
Moreover, we decompose the torsion 2-form $T_\mu$ and introduce the
components of the Hodge dual of the torsion $^\star T^\alpha={\check
  T}_{\beta\gamma}{}^{\alpha}\,\vartheta^{\beta\gamma}/2$. Then the
constitutive tensor of spacetime becomes
\begin{equation}\label{xxx1}
  \kappa_{\gamma\delta}{} ^{\alpha\beta}= \lambda_0\left(
  \eta^{\alpha\beta} {}_{\gamma\delta}-2\ell^2\, {\check
    T}^{\alpha\beta}{}_{\mu}\,T_{\gamma\delta}{}^\mu\right).
\end{equation}
We can raise the two first indices with the $\eta$ according to the
definition (see \cite{Birk})
\begin{equation}\label{xxx2}
  \chi^{\gamma\delta\alpha\beta}=-\,\frac{\sqrt{-g}}{2}\,\eta^{
    \gamma\delta \lambda\nu }\, \kappa_{\lambda\nu}{} ^{\alpha\beta}\,.
\end{equation}
Then, after some algebra, we find
\begin{equation}\label{xxx6}
  \chi^{\alpha\beta\gamma\delta} = 2\lambda_0\,\sqrt{-g}\left(
    g^{\alpha[\gamma}\,g^{\delta]\beta} +\ell^2\, {\check
      T}^{\alpha\beta}{}_{\mu}\, {\check
      T}^{\gamma\delta\mu}\,\right)\,.
\end{equation}

Because of the symmetries 
\begin{equation}\label{xxx7}
  \chi^{\alpha\beta\gamma\delta} = \chi^{\gamma\delta\alpha\beta} =-
  \chi^{\beta\alpha\gamma\delta} = -\chi^{\alpha\beta\delta\gamma}\,,
\end{equation}
the tensor density $\chi$ has 21 independent components. Its totally
antisymmetric (axion) piece
\begin{equation}\label{xxx8}
\chi^{[\alpha\beta\gamma\delta]} =\frac{\lambda_0\ell^2}{6}\,\sqrt{-g}\,\eta
{}^{\alpha\beta\gamma\delta}\,{\check T}^{\mu\nu\rho}\, T_{\mu\nu\rho}\,,
\end{equation} 
carries 1 independent component. It is known (see \cite{Birk}) that the 
axion does {\em not\/} affect a path of light (although the polarization 
vector feels the axion, see \cite{clausannalen}). Hence the principal 
piece $\chi^{\alpha\beta\gamma\delta} - \chi^{[\alpha\beta\gamma\delta]}$
alone with its 20 independent components determines the light cone structure.

\subsection{Derivation of the double light cone structure}

After $\chi$ is known explicitly, see (\ref{xxx6}), we determine the 
corresponding TR-tensor density\footnote{The name Tamm-Rubilar tensor 
density was suggested in \cite{Birk}, see the references given there.} 
$\cal G$. In holonomic coordinates (natural frames), we have 
\begin{equation}\label{GU8}
 {\cal G}^{ijkl}:=\frac{1}{4!}\,\hat{\epsilon}_{mnpq}\,
 \hat{\epsilon}_{rstu}\,\chi^{\,mnr(i}\, \chi^{\,j|ps|k}\, \chi^{\,l)qtu }, 
\qquad {\cal G}^{ijkl}={\cal G}^{(ijkl)}
\end{equation}
(see \cite{Birk} and \cite{drguillermo}, for example; in the framework
of general relativity, Perlick \cite{perlick} describes related methods).

For the O(3)-symmetric torsion of (\ref{torsion10}), we now choose an
orthonormal frame with the metric $g_{\alpha\beta} = {\rm diag}(+1,-1,-1,-1)$.
Application of the Hodge star to (\ref{torsion10}) yields
\begin{equation}\label{torsion10'}
  ^\star T^{\alpha}|_{{\rm O}(3)} =\pmatrix{ 
 { T}_{\hat 0\hat 1}{}^{\hat 0} \,^\star{\vartheta}^{\hat 0\hat 1} \cr 
 { T}_{\hat 0\hat 1}{}^{\hat 1} \,^\star{\vartheta}^{\hat 0\hat 1} \cr 
 { T}_{\hat 0\hat 2}{}^{\hat 2} \,^\star{\vartheta}^{\hat 0\hat 2}
+{ T}_{\hat 1\hat 2}{}^{\hat 2} \,^\star{\vartheta}^{\hat 1\hat 2} \cr 
 { T}_{\hat 0\hat 3}{}^{\hat 3} \,^\star{\vartheta}^{\hat 0\hat 3}   
+{ T}_{\hat 3\hat 1}{}^{\hat 3} \,^\star{\vartheta}^{\hat 3\hat 1}}=
{\frac 1 \ell}\pmatrix{ -f \,{\vartheta}^{\hat 2\hat 3} \cr 
h\,{\vartheta}^{\hat 2\hat 3} \cr  k \,{\vartheta}^{\hat 3\hat 1}
+g \,{\vartheta}^{\hat 0\hat 3} \cr  k \,{\vartheta}^{\hat 1\hat 2}   
-g \,{\vartheta}^{\hat 0\hat 2}} 
\end{equation} 
(recall that $\eta^{\hat 0\hat 1\hat 2\hat 3}=-1/\sqrt{-g}$).  
We read off the following components:
\begin{eqnarray}\label{xxx9}
&&{\check T}_{\hat 2\hat 3}{}^{\hat 0}=-f/\ell\,, 
\quad {\check T}_{\hat 2\hat 3}{}^{\hat 1}= h/\ell\,, 
\quad {\check T}_{\hat 3\hat 1}{}^{\hat 2}= k/\ell \,,\nonumber\\
&& {\check T}_{\hat 0\hat 3}{}^{\hat 2}= g/\ell  \,, 
\quad {\check T}_{\hat 1\hat 2}{}^{\hat 3}= k/\ell  \,, 
\quad {\check T}_{\hat 0\hat 2}{}^{\hat 3}=-g/\ell \,.
\end{eqnarray} 

We substitute (\ref{xxx9}) into (\ref{xxx6}):
\begin{eqnarray}
\chi^{\hat 0\hat 1\hat 0\hat 1}&=&-\lambda_0\,, \quad 
\chi^{\hat 0\hat 2\hat 0\hat 2}=
\chi^{\hat 0\hat 3\hat 0\hat 3}= - \lambda_0(1 + 2g^2)\,,\nonumber\\
\chi^{\hat 2\hat 3\hat 2\hat 3}&=&\lambda_0[1 + 2(f^2-h^2)]\,, \quad 
\chi^{\hat 3\hat 1\hat 3\hat 1}=
\chi^{\hat 1\hat 2\hat 1\hat 2}=\lambda_0(1 - 2k^2)\,, \nonumber \\
\chi^{\hat 0\hat 2\hat 1\hat 2}&=& \chi^{\hat 1\hat 2\hat 0\hat 2} =
-\chi^{\hat 0\hat 3\hat 3\hat 1}= -\chi^{\hat 3\hat 1\hat 0\hat 3} =
-2\lambda_0g\,k\,.\label{GU12}
\end{eqnarray} 
Obviously, the axion part (\ref{xxx8}) vanishes for the spherically
symmetric torsion: $\chi^{[\alpha\beta\gamma\delta]} = 0$. 
In $6\times 6$ form (the corresponding matrix is symmetric), we have
(with the bivector indices $I,J=1,\dots,6$ defined as $1=\hat 0\hat 1,
2=\hat 0\hat 2, 3=\hat 0\hat 3, 4=\hat 2\hat 3, 5=\hat 3\hat 1, 
6=\hat 1\hat 2$)
\begin{equation}\label{nonvan2}
  {\chi}^{IK} = \pmatrix{B & D \cr C & A},
\end{equation}
where the $3\times 3$ blocks read:
\begin{eqnarray}
A &=& -\,\lambda_0\pmatrix{1&0&0\cr 0&1 + 2g^2&0\cr 0&0& 1 + 2g^2},\quad
C = \lambda_0\pmatrix{0&0&0\cr 0&0&-2gk\cr 0&2gk&0},\\
B &=&\ \lambda_0\pmatrix{1 + 2(f^2 - h^2)&0&0\cr 0&1 - 2k^2&0\cr 0&0&1 - 2k^2},
\quad D = C^{\rm T}.
\end{eqnarray}

The corresponding Fresnel equation determines the components of the wave 
covector (1-form) $q_i$ of the propagating wave:
\begin{equation} \label{Fresnel}
{\cal G}^{ijkl}q_i q_j q_k q_l = 0 \,.
\end{equation}
Explicitly, it is found in this case to be 
\begin{eqnarray}
-\lambda_0^3 && \left[ (1 + 2g^2) q_{\hat 0}^2 + 4gk\,q_{\hat 0}\,q_{\hat 1}
  -(1 - 2k^2) q_{\hat 1}^2 - [1 + 2(g^2-k^2)](q_{\hat 2}^2 + q_{\hat 3}^2)
\right]\times\nonumber \\ 
  && \left[ (1 + 2g^2)q_{\hat 0}^2 + 4gk\,q_{\hat 0}\,q_{\hat 1} 
  -(1 - 2k^2)q_{\hat 1}^2 -[1 + 2(f^2-h^2)](q_{\hat 2}^2 + q_{\hat 3}^2)
\right] =0.\label{GU14}
\end{eqnarray}
This result was also checked with the help of computer algebra. 
Thus, we verify that in this case we have birefringe, i.e., a double
lightcone structure. The respective {\it optical} metrics are, up to 
a conformal factor, given by
\begin{equation}\label{GU15}
g_{(1)}^{\alpha\beta}=\left(\begin{array}{cccc} 
1 + 2g^2 & 2gk  & 0  & 0 \\ 2gk & - 1 + 2k^2 & 0 & 0 \\        
0  &  0    & - 1 + 2(k^2-g^2) & 0\\ 
0  &  0   & 0  & - 1 + 2(k^2-g^2)\end{array}\right),
\end{equation} 
\begin{equation}\label{GU16}
  g_{(2)}^{\alpha\beta}=\left(\begin{array}{cccc}           
1 + 2 g^2 & 2 gk  & 0  & 0 \\ 2gk  & - 1 + 2k^2 &  0  & 0 \\ 
0  &  0  & - 1 + 2(h^2-f^2) & 0 \\ 
0  &  0  &   0   & - 1 + 2(h^2-f^2)\end{array}\right).
\end{equation}

\section{Torsion in Friedmann cosmology}

As another example, we consider light propagation on the background
of torsion in a Friedmann universe. The isotropy group is SO(3).
The torsion pieces left over after we impose the homogeneity and isotropy
conditions read, with $u=u(t)$ and $v=v(t)$ (see
\cite{nara}, \cite{goenner}, and \cite{albert}, e.g.):
\begin{equation}\label{torsion10fried}
  T^{\alpha}|_{{\rm F}} =\pmatrix{ 0\cr T_{\hat 0\hat
      1}{}^{\hat 1} \,{\vartheta}^{\hat 0\hat 1}+ T_{\hat 2\hat
      3}{}^{\hat 1} \,{\vartheta}^{\hat 2\hat 3} \cr T_{\hat 0\hat
      2}{}^{\hat 2} \,{\vartheta}^{\hat 0\hat 2}+ T_{\hat 3\hat
      1}{}^{\hat 2} \,{\vartheta}^{\hat 3\hat 1} \cr T_{\hat 0\hat
      3}{}^{\hat 3} \,{\vartheta}^{\hat 0\hat 3}+ T_{\hat 1\hat
      2}{}^{\hat 3} \,{\vartheta}^{\hat 1\hat 2} }=
   {\frac 1 \ell}\pmatrix{ 0\cr u
    \,{\vartheta}^{\hat 0\hat 1}+ v \,{\vartheta}^{\hat 2\hat 3} \cr u
    \,{\vartheta}^{\hat 0\hat 2}+ v \,{\vartheta}^{\hat 3\hat 1} \cr u
    \,{\vartheta}^{\hat 0\hat 3}+ v \,{\vartheta}^{\hat 1\hat 2}}.
\end{equation} 
By simple algebra we find
\begin{equation}
u^2=\frac{\ell^2}{9}T_{\alpha\beta}{}^\beta\,T^{\alpha\gamma}{}_\gamma\,,\qquad
v^2=\frac{\ell^2}{6}T_{[\alpha\beta\gamma]}\,T^{[\alpha\beta\gamma]}\,.
\end{equation}
The dual of (\ref{torsion10fried}),
\begin{equation}\label{torsion10fried1}
  ^\star T^{\alpha}|_{{\rm F}} = {\frac 1 \ell}\pmatrix{ 0\cr v 
   \,{\vartheta}^{\hat 0\hat 1}-u \,{\vartheta}^{\hat 2\hat 3} \cr v
    \,{\vartheta}^{\hat 0\hat 2}- u \,{\vartheta}^{\hat 3\hat 1} \cr v
    \,{\vartheta}^{\hat 0\hat 3}- u \,{\vartheta}^{\hat 1\hat 2}
    } \,,
\end{equation} 
allows us to read off the components of the dual of the torsion:
\begin{equation}\label{torsion10fried2} 
  {\check T}_{\hat 0\hat 1}{}^{\hat 1}={\check T}_{\hat 0\hat
    2}{}^{\hat 2}={\check T}_{\hat 0\hat 3}{}^{\hat
    3}= v(t)/\ell\,,\quad {\check T}_{\hat 2\hat 3}{}^{\hat
    1}={\check T}_{\hat 3\hat 1}{}^{\hat 2}={\check T}_{\hat 1\hat
    2}{}^{\hat 3}=-u(t)/\ell. 
\end{equation}
We substitute this result into (\ref{xxx6}):
\begin{eqnarray}
  && \chi^{\hat 0\hat 1\hat 0\hat 1} =\chi^{\hat 0\hat 2\hat 0\hat 2}
   =\chi^{\hat 0\hat 3\hat 0\hat 3 }= -\lambda_0(1 + 2v^2)\,, \\ 
  && \chi^{\hat 2\hat 3\hat 2\hat 3} =\chi^{\hat 3\hat 1\hat 3\hat 1}
   =\chi^{\hat 1\hat 2\hat 1\hat 2 }=
    \lambda_0(1 - 2u^2)\,, \\ && \chi^{\hat 0\hat 1\hat 2\hat 3}
    =\chi^{\hat 0\hat 2\hat 3\hat 1 }=\chi^{\hat 0\hat 3\hat 1\hat 2}
    = \chi^{\hat 2\hat 3\hat 0\hat 1}=\chi^{\hat 3\hat 1\hat 0\hat 2}
    =\chi^{\hat 1\hat 2\hat 0\hat 3}=-2\lambda_0\,uv\,.
\end{eqnarray}
In contrast to the spherically symmetric case, here the axion part 
(\ref{xxx8}) is nontrivial: $\chi^{[\alpha\beta\gamma\delta]} = \alpha
\,\epsilon^{\alpha\beta\gamma\delta}$ with the axion field $\alpha = 
-2\lambda_0\,uv$. In the $6\times 6$ matrix form (\ref{nonvan2}), the 
$3\times 3$ constitutive matrices now read:
\begin{equation}
A^{ab} = -\,\lambda_0(1 + 2v^2)\delta^{ab},\quad
B_{ab} = \lambda_0(1 - 2u^2)\delta_{ab},\quad
C^a{}_b = D_b{}^a = -\,2\lambda_0uv\delta^a_b.
\end{equation}

As above, we calculate the TR-tensor density $\cal G$ and check this 
with the help of computer algebra. The corresponding Fresnel equation reads:
\begin{equation}
  -\lambda_0^3(1 + 2v^2)\left[(1 + 2v^2)q_{\hat 0}^2 -(1 - 2u^2)
  (q_{\hat 1}^2 + q_{\hat 2}^2 + q_{\hat 3}^2) \right]^2 =0\,.
\end{equation}
Perhaps surprizingly we find a single effective light cone: no
birefringence, just a different speed of light as compared to the case
without torsion.  Up to a conformal factor, the optical metric is obviously
\begin{equation}\label{opt}
g_{\rm opt}^{\alpha\beta}=\left(\begin{array}{cccc}  
1 + 2v^2 & 0 &  0  & 0 \\ 0 & -1 + 2u^2 &  0  & 0 \\ 
0  &  0  & -1 + 2u^2 & 0 \\ 0 & 0 &  0  & -1 + 2u^2\end{array}\right).
\end{equation} 
This result (reduction of the quartic Fresnel surface to a unique light
cone) implies that we can rewrite the constitutive tensor in the
conventional form:
\begin{equation}
\chi^{\alpha\beta\gamma\delta} = \varphi\,\sqrt{-g_{\rm opt}}\left(
g_{\rm opt}^{\alpha\gamma}\,g_{\rm opt}^{\beta\delta} - 
g_{\rm opt}^{\alpha\delta}\,g_{\rm opt}^{\beta\gamma}\right) 
+ \alpha\,\epsilon^{\alpha\beta\gamma\delta}
\end{equation}
with the effective dilaton field $\varphi = \lambda_0\sqrt{(1 + 2v^2)
(1 - 2u^2)}$. 

According to the metric (\ref{opt}), photons would propagate isotropically 
but with a torsion-dependent speed. In astrophysical observations, this
could show up in a certain deviation from the cosmological redshift
predictions of general relativity theory. When combined with additional
studies on the behavior of particles and fields with spin, such observations
may provide an experimental proof for the existence of torsion.

\bigskip
{\bf Acknowledgments}. YNO's work was supported by the Deutsche 
Forschungsgemeinschaft (Bonn), project HE~528/20-1. The authors are 
grateful to Christian Heinicke (Cologne) for drawing our attention to 
the talk of O.~Preuss and to O.~Preuss for sending us a copy of his thesis. 
We appreciate useful remarks by Volker Perlick (Cologne/Berlin) on 
Friedmannian cosmology.

\end{document}